\documentclass[a4paper, runningheads]{llncs}

\usepackage[T1]{fontenc}
\usepackage[utf8]{inputenc}

\usepackage{microtype}
\usepackage[hidelinks]{hyperref}
\usepackage{tabu}

\usepackage{url}

\begin{document}

\mainmatter              

\title{Discovering and Proving Invariants\\in Answer Set Programming and Planning}
\titlerunning{Discovering and Proving Invariants in ASP and Planning}
\author{Patrick Lühne}
\authorrunning{Patrick Lühne}
\tocauthor{Patrick Lühne}
\institute{University of Potsdam, Germany, \\
	\email{patrick.luehne@cs.uni-potsdam.de}}

\maketitle

\begin{abstract}
Answer set programming (ASP) and planning are two widely used paradigms for solving logic programs with declarative programming.
In both cases, the quality of the input programs has a major influence on the quality and performance of the solving or planning process.
Hence, programmers need to understand how to make their programs efficient and still correct.
In my PhD studies, I explore how input programs can be improved and verified automatically as a means to support programmers.
One of my research directions consists in discovering invariants in planning programs without human support, which I implemented in a system called \textit{ginkgo}.
Studying dynamic systems in greater depth, I then developed \textit{plasp}~3 with members of my research group, which is a significant step forward in effective planning in ASP.
As a second research direction, I am concerned with automating the verification of ASP programs against formal specifications.
For this joint work with Lifschitz’s group at the University of Texas, I developed a verification system called \textit{anthem}.
In my future PhD studies, I will extend my research concerning the discovery and verification of ASP and planning problems.
\end{abstract}

\section{Introduction}

Answer set programming (ASP) and planning are two widely used paradigms for solving logic programs with declarative programming.
While ASP aims to be as general-purpose as possible, planning focuses on dynamic systems, where sequences of actions are searched in order to achieve specific goals.
As with other knowledge representation paradigms, quality and performance in solving and planning not only depend on the implementations of solvers and planners but also on the quality of the input program specifications.
From the perspective of programmers working with solvers and planners, it is, hence, fundamentally important to understand whether their programs are efficient and, more importantly, correct with respect to their specifications.


In my PhD studies, I explore how input programs can be improved and verified automatically as a means to support programmers.
Both of these objectives relate to \emph{invariants} of logic programs---properties that preserve the correctness of a program.

My first research direction consists in \emph{discovering} invariants without human support.
For this purpose, I developed the system \textit{ginkgo}, which continuously discovers invariants in planning problems by generalizing conflict constraints learned by an ASP solver (Section~\ref{sec:ginkgo}).
While working with planning problems in the \emph{planning domain definition language} (PDDL~\cite{mcdermott98a}), I further implemented \textit{plasp}~3, the third generation of an ASP planning system.
With \textit{plasp}, PDDL programs can be solved with established ASP solvers such as \textit{clingo}.
Based on this effort, members of our research group and I showed how to make planning in ASP more effective with parallel planning (Section~\ref{sec:plasp}).

As a second research direction, I investigate how to automate the \emph{verification} of ASP programs against formal specifications in a subset of the input language of \textit{clingo} in cooperation with Vladimir Lifschitz’s group at the University of Texas.
\textit{anthem}, another system that I developed, performs this task by translating ASP programs to first-order logic formulas to validate the strong equivalence against a specification by a theorem prover (Section~\ref{sec:anthem}).

In future work, I will extend my research concerning the discovery and verification of ASP and planning problems.
Section~\ref{sec:future-work} discusses such directions, before Section~\ref{sec:conclusions} concludes this extended abstract.

\section{\textit{ginkgo}—Discovering Invariants in ASP Planning}
\label{sec:ginkgo}

\emph{Conflict learning} has become a base technology in Boolean constraint solving, and, in particular, answer set programming.
However, learned constraints are only valid for a currently solved problem instance and do not carry over to similar instances.
To address this issue, I developed a framework featuring an integrated feedback loop that allows for reusing conflict constraints (published in Technical Communications of ICLP~2016~\cite{gekakalurosc16a}).
The idea is to extract (propositional) conflict constraints, generalize and validate them, and reuse them as integrity constraints.
In this way, an input program is continuously extended with automatically discovered invariants.
Although I explored this approach in the context of dynamic systems (specifically, PDDL planning), the ultimate objective is to overcome the issue that learned knowledge is bound to specific problem instances.

I implemented this workflow in two systems, namely, a variant of the ASP solver \textit{clasp} that extracts integrity constraints, along with the downstream system \textit{ginkgo}\footnote{\url{https://github.com/potassco/ginkgo}} for generalizing and validating them.
\textit{ginkgo} finds invariants by first deriving candidate properties (learned constraints that are generalized over the temporal domain).
These properties are then checked for invariance.
This relies on automated proofs that I fully implemented in ASP with meta encodings.

\section{\textit{plasp}~3—Towards Effective ASP Planning}
\label{sec:plasp}

Emerging from my work with ASP-based planning in the \textit{ginkgo} system, I implemented the third installment of \textit{plasp}\footnote{\url{https://github.com/potassco/plasp}}.
While earlier versions of \textit{plasp} were pure PDDL-to-ASP translators~\cite{gekaknsc11a}, \textit{plasp}~3 was conceived to provide a flexible platform to experiment with a variety of techniques to make planning in ASP more effective (published at LPNMR 2017~\cite{digelu17a} and TPLP~2019~(3)~\cite{digelu19a}).

For this purpose, I reimplemented \textit{plasp}, while widening the range of accepted PDDL features in comparison to the previous versions.
Further, our research group developed novel planning encodings, some inspired by SAT planning and others exploiting ASP features such as well-foundedness.
I designed \textit{plasp}~3 such that it handles multivalued fluents and, hence, captures both PDDL as well as SAS planning formats.
Third, enabled by multishot ASP solving, advanced planning algorithms are offered, also borrowed from SAT planning.
Empirical analyses show that these techniques have a significant impact on the performance of ASP planning.

\section{\textit{anthem}—Verifying the Correctness of ASP Programs}
\label{sec:anthem}

Harrison, Lifschitz, and Raju have extended the definition of program completion to a subset of the input language of \textit{clingo}~\cite{har17a,lif16}.
The aim of their work is to extend the applicability of formal verification methods to ASP by turning logic programs into completed definitions.
This can also be understood as a translation from \textit{clingo}’s input language to first-order logic formulas.

In cooperation with their research group at the University of Texas, I developed a system called \textit{anthem},\footnote{\url{https://github.com/potassco/anthem}} which performs the completion of logic programs automatically (published in ASPOCP 2018~\cite{lilusc18a}).
After translating and simplifying formulas with \textit{anthem}, programmers can see more clearly what exactly their program solves.

Furthermore, the first-order logic representation can be used to verify the strong equivalence of logic programs with a computer-assisted proof.
The proof can then be conducted by a theorem prover, popuplar examples for which include \emph{E}~\cite{Schulz:LPAR-2013}, \emph{Coq}~\cite{coq}, and \emph{Prover9}~\cite{prover9}.
To that end, I extended \textit{anthem} to perform a new translation $\tau^*$ from logic programs to a \emph{finite} set of first-order sentences.
With this translation, \textit{anthem} and the theorem prover \textit{Vampire}~\cite{kovvor13a} can be used in conjunction to programmatically verify the strong equivalence of two positive input programs (to be published at LPNMR~2019~\cite{lilusc19a}).

My next goal is to extend \textit{anthem} such that it can be used to quickly test whether ASP programs fulfill given invariants.
This can be achieved by combining the previously implemented completion step and translation $\tau^*$.
With such a tool, programmers could start writing programs by first making a formal specification, against which their code is later verified.

\section{Future Work}
\label{sec:future-work}

As stated before, my most recent work focuses on using theorem provers to verify that logic programs comply with a given specification.
This indirection of proving invariants through first-order logic might turn out particularly useful when coming back to my earlier research on the \textit{ginkgo} system.
This is because there are many established first-order theorem provers, which might make for a stronger proof system than the counterexample-based validation method currently employed by \textit{ginkgo}.

Furthermore, I will expand my research in the field of PDDL planning.
Building on \textit{plasp}~3, I want to explore how far planning in ASP can be pushed, with the objective of achieving performance on par with state-of-the-art SAT planners such as \emph{Madagascar}~\cite{rintanen14a}.

Finally, I am always exploring opportunities to apply the planning-related techniques to the broader scope of general ASP programs.
This involves the research areas of automatic modeling, program synthesis, and superoptimization, which I want to further familiarize myself in my upcoming PhD studies.

\section{Conclusions}
\label{sec:conclusions}

Concerning the automatic discovery and verification of invariants in logic programs, I already made insightful progress.
In my early PhD projects, 
I showed the feasibility of reusing learned conflict constraints in ASP planning by means of generalization with my \textit{ginkgo} system.
I further studied dynamic systems as such and helped making ASP planning much more effective with \textit{plasp}~3 and performance-wise closer to state-of-the-art SAT planners than previous attempts.
I believe that these two systems I developed could make use of refined invariant finding techniques, which is one of the things I want to study in the remainder of my PhD studies.

Furthermore, I am researching automated verification techniques in multiple contexts.
First, as a means to validate potential candidate invariants within \textit{ginkgo}.
Second, to completely automate the process of testing ASP programs against formal specifications, which is the objective of my currently work-in-progress system \textit{anthem}.
This is a technique that could later be useful for other parts of my research as well.

To my mind, there are many interesting aspects of discovering and verifying invariants ahead that I want to address in my PhD studies.
This also includes more practical applications such as making ASP planning yet more effective.

\bibliography{paper}

\begin{thebibliography}{10}

\bibitem{coq}
Bruno Barras, Samuel Boutin, Cristina Cornes, Judica{\"e}l Courant,
  Jean-Christophe Filliatre, Eduardo Gimenez, Hugo Herbelin, Gerard Huet, Cesar
  Munoz, Chetan Murthy, et~al.
\newblock The {C}oq proof assistant reference manual: Version 8.6.
\newblock \url{https://coq.inria.fr/distrib/current/refman/}, 2016.

\bibitem{digelu17a}
Y.~Dimopoulos, M.~Gebser, P.~Lühne, J.~Romero, and T.~Schaub.
\newblock \textit{plasp}~3: Towards effective {ASP} planning.
\newblock In {\em Proceedings of the Fourteenth International Conference on
  Logic Programming and Nonmonotonic Reasoning (LPNMR'17)}, pages 286--300.
  Springer-Verlag, 2017.

\bibitem{digelu19a}
Y.~Dimopoulos, M.~Gebser, P.~Lühne, J.~Romero, and T.~Schaub.
\newblock {p}lasp 3: Towards effective {ASP} planning.
\newblock {\em Theory and Practice of Logic Programming}, 19(3):477–504,
  2019.

\bibitem{gekakalurosc16a}
M.~Gebser, R.~Kaminski, B.~Kaufmann, P.~Lühne, J.~Romero, and T.~Schaub.
\newblock Answer set solving with generalized learned constraints.
\newblock In M.~Carro and A.~King, editors, {\em Technical Communications of
  the Thirty-second International Conference on Logic Programming (ICLP'16)},
  volume~52, pages 9:1--9:15. Open Access Series in Informatics (OASIcs), 2016.

\bibitem{gekaknsc11a}
M.~Gebser, R.~Kaminski, M.~Knecht, and T.~Schaub.
\newblock plasp: A prototype for {PDDL}-based planning in {ASP}.
\newblock In J.~Delgrande and W.~Faber, editors, {\em Proceedings of the
  Eleventh International Conference on Logic Programming and Nonmonotonic
  Reasoning (LPNMR'11)}, volume 6645 of {\em Lecture Notes in Artificial
  Intelligence}, pages 358--363. Springer-Verlag, 2011.

\bibitem{har17a}
Amelia Harrison, Vladimir Lifschitz, and Dhananjay Raju.
\newblock Program completion in the input language of gringo.
\newblock Submitted for publication, 2017.

\bibitem{kovvor13a}
L.~Kov{\'a}cs and A.~Voronkov.
\newblock First-order theorem proving and {V}ampire.
\newblock In N.~Sharygina and H.~Veith, editors, {\em Proceedings of the
  Twenty-fifth International Conference on Computer Aided Verification
  (CAV'13)}, volume 8044, pages 1--35. Springer-Verlag, 2013.

\bibitem{lilusc18a}
V.~Lifschitz, P.~L{\"u}hne, and T.~Schaub.
\newblock anthem: Transforming gringo programs into first-order theories
  (preliminary report).
\newblock In J.~Fandinno and J.~Fichte, editors, {\em Proceedings of the
  Eleventh Workshop on Answer Set Programming and Other Computing Paradigms
  (ASPOCP'18)}, 2018.

\bibitem{lilusc19a}
V.~Lifschitz, P.~Lühne, and T.~Schaub.
\newblock Verifying strong equivalence of programs in the input language of
  \textsc{gringo} \emph{(to be published)}.
\newblock In {\em Proceedings of the Fifteenth International Conference on
  Logic Programming and Nonmonotonic Reasoning (LPNMR'19)}. Springer-Verlag,
  2017.

\bibitem{lif16}
Vladimir Lifschitz.
\newblock Achievements in answer set programming (preliminary report).
\newblock In {\em Working Notes of the Workshop on Answer Set Programming and
  Other Computing Paradigms}, 2016.

\bibitem{prover9}
W.~McCune.
\newblock Prover9 and {M}ace4.
\newblock \url{http://www.cs.unm.edu/~mccune/prover9/}, 2005--2010.

\bibitem{mcdermott98a}
D.~McDermott.
\newblock {PDDL} --- the planning domain definition language.
\newblock Technical Report CVC TR-98-003/DCS TR-1165, Yale Center for
  Computational Vision and Control, 1998.

\bibitem{rintanen14a}
J.~Rintanen.
\newblock Madagascar: Scalable planning with {SAT}.
\newblock In M.~Vallati, L.~Chrpa, and T.~McCluskey, editors, {\em Proceedings
  of the Eighth International Planning Competition (IPC'14)}, pages 66--70.
  University of Huddersfield, 2014.

\bibitem{Schulz:LPAR-2013}
Stephan Schulz.
\newblock {System Description: E~1.8}.
\newblock In Ken McMillan, Aart Middeldorp, and Andrei Voronkov, editors, {\em
  Proc.\ of the 19th LPAR, Stellenbosch}, volume 8312 of {\em LNCS}. Springer,
  2013.

\end{thebibliography}

\end{document}